\documentclass[manuscript]{aastex6}
\usepackage[utf8]{inputenc}
\usepackage{wasysym}
\usepackage{afterpage}
\accepted{02/03/2017}

\begin{document}

\title{The shadow knows: using shadows to investigate the structure of the pretransitional disk of HD 100453}

\author{Zachary C. Long\altaffilmark{1}, Rachel B. Fernandes\altaffilmark{1}, Michael Sitko\altaffilmark{1,2}, Kevin Wagner\altaffilmark{3}, Takayuki Muto\altaffilmark{4}, Jun Hashimoto\altaffilmark{5}, Katherine Follette\altaffilmark{6,7}, Carol A. Grady\altaffilmark{8}, Misato Fukagawa\altaffilmark{9}, Yasuhiro Hasegawa\altaffilmark{10,11,12}, Jacques Kluska\altaffilmark{13}, Stefan Kraus\altaffilmark{13}, Satoshi Mayama\altaffilmark{14}, Michael W. McElwain\altaffilmark{15},  Daehyeon Oh\altaffilmark{5,14}, Motohide Tamura\altaffilmark{5,16,17}, Taichi Uyama\altaffilmark{16}, John P. Wisniewski\altaffilmark{18}, Yi Yang\altaffilmark{5,14}}
\altaffiltext{1}{Department of Physics, University of Cincinnati, Cincinnati, OH 45221, USA}
\altaffiltext{2}{Center for Extrasolar Planetary Studies, Space Science Institute,  4750 Walnut St, Suite 205, Boulder, CO 80301}
\altaffiltext{3}{Department of Astronomy/Steward Observatory, The University of Arizona, 933 N. Cherry Avenue, Tucson, AZ 85721}
\altaffiltext{4}{Division of Liberal Arts, Kogakuin University, 1-24-2 Nishi-Shinjuku, Shinjuku-ku, Tokyo, 163-8677, Japan}
\altaffiltext{5}{National Astronomical Observatory of Japan, 2-21-1, Osawa, Mitaka, Tokyo, 181-8588, Japan}
\altaffiltext{6}{Kavli Institute for Particle Astrophysics and Cosmology, Stanford University, Stanford, California 94305, USA}
\altaffiltext{7}{NASA Sagan Fellow}
\altaffiltext{8}{Eureka Scientific, 2452 Delmer St. Suite 100, Oakland CA 96402, USA}
\altaffiltext{9}{Division of Particle and Astrophysical Science, Graduate School of Science, Nagoya University, Nagoya, Japan}
\altaffiltext{10}{Jet Propulsion Laboratory, California Institute of Technology, Pasadena, CA 91109, USA}
\altaffiltext{11}{Institute of Astronomy and Astrophysics, Academia Sinica (ASIAA), Taipei 10641, Taiwan}
\altaffiltext{12}{Division of Theoretical Astronomy, National Astronomical Observatory of Japan, Osawa, Mitaka, Tokyo 181-8588, Japan}
\altaffiltext{13}{University of Exeter Astrophysics Group, School of Physics, Stocker Road, Exeter, Devon EX4 4QL UK}
\altaffiltext{14}{Department of Astronomical Science, The Graduate University for Advanced Studies (SOKENDAI), Shonan Village, Hayama, Kanagawa 240-0193 Japan}
\altaffiltext{15}{Exoplanets and Stellar Astrophysics Laboratory, Code 667, NASA's Goddard Space Flight Center, Greenbelt, MD 20771, USA}
\altaffiltext{16}{Department of Astronomy and RESCUE, The University of Tokyo, 7-3-1, Hongo, Bunkyo-ku, Tokyo, 113-0033, Japan}
\altaffiltext{17}{Astrobiology Center of NINS, 2-21-1, Osawa, Mitaka, Tokyo, 181-8588, Japan}
\altaffiltext{18}{Homer L. Dodge Department of Physics, University of Oklahoma, Norman, OK 73071, USA}

\begin{abstract}
We present GPI polarized intensity imagery of HD 100453 in Y-, J-, and K1 bands which reveals an inner gap ($9 - 18$ au), an outer disk ($18-39$ au) with two prominent spiral arms, and two azimuthally-localized dark features also present in SPHERE total intensity images \citep{wagner15}. SED fitting further suggests the radial gap extends to $1$ au. The narrow, wedge-like shape of the dark features appears similar to predictions of shadows cast by a inner disk which is misaligned with respect to the outer disk. Using the Monte Carlo radiative transfer code HOCHUNCK3D \citep{whitney13}, we construct a model of the disk which allows us to determine its physical properties in more detail. From the angular separation of the features we measure the difference in inclination between the disks (45$\degr$), and their major axes, PA = 140$\degr$ east of north for the outer disk and 100$\degr$ for the inner disk. We find an outer disk inclination of $25 \pm 10\degr$ from face-on in broad agreement with the \citet{wagner15} measurement of 34$\degr$. SPHERE data in J- and H-bands indicate a reddish disk which points to HD 100453 evolving into a young debris disk.

\end{abstract}

\section{INTRODUCTION}

Images of SAO 206462 \citep{stolker16} and HD 142527 \citep{marino15} have revealed azimuthally-localized dark features in their outer disks. Both studies interpret the features as shadows cast by an optically thick, non-coplanar inner disk. Such an inner disk may be indicative of the existence of planets or large dynamical changes in the disk's history. Modeling of the disk structures can provide a predictive tool for where low mass companions may be hiding. Sequential follow-up observations can then detect these worlds, as was the case with the young gas giant $\beta$ Pic b, which due to its inclined orbit is driving a similar gravitational warp in the inner disk that was seen before the planet \citep{apai15}.

Another disk that exhibits azimuthally-localized dark features is associated with HD 100453A (A9V, luminosity L $\sim$ 9 L$_{\astrosun}$, mass M $\sim$ 1.7 M$_{\astrosun}$ \citep{dominik03}, d = $103 \pm 3$ pc \citep{gaia16}) which has an M4 companion at a separation of $1\farcs06 \pm 0\farcs02$ \citep{chen06}.  Total intensity (TI) imaging at Y-K2 bands, obtained by \citet{wagner15} with the extreme AO imager Spectro-Polarimetric High-contrast Exoplanet REsearch (SPHERE), revealed two spiral arms, dark features, and the outer extent of the large radial gap inferred from the IR spectral energy distribution (SED) \citep{khalafinejad16,maaskant13}.  

We present Gemini Planet Imager (GPI) polarized intensity (PI) imagery \citep{macintosh14} of HD 100453. This dataset, the SPHERE data, and the infrared SED are modeled using the 3D Monte Carlo radiative transfer code HOCHUNK3D \citep{whitney13}. This code allows for two, independent and radially separated two-layer disks which need not be coplanar, and the use of multiple grain opacity models. Model images are compared to image data, while simultaneously fitting the SED, to find the cause of the dark features and investigate how structural changes within the disk affect their shape and locations. 

\section{OBSERVATIONS AND DATA REDUCTION}

\subsection{GPI Observations and Data Reduction}

HD 100453 was observed with GPI on 2015 April 10-11th UT using polarimetric differential imaging mode in J-, Y-, and K1-bands. GPI’s “direct” mode utilizes polarization to suppress stellar light, but not a coronagraphic mask, thereby sacrificing contrast in favor of a tighter inner working angle. We used the shortest exposure time (1.49 seconds) to minimize saturation and co-added 10 frames to increase the signal-to-noise ratio. The half-wave plate angle was rotated from 0 to 67.5$\degr$ with 22.5$\degr$ steps to obtain linear polarization. This sequence was repeated 41, 35, and 36 times, resulting in total exposure time of 41, 35, and 36 minutes for Y-, J-, and K1-bands, respectively. The averaged airmass values were 1.21 and 1.19 in the $J$- and $K1$-bands, respectively.

The data were reduced using the standard polarized data reduction recipe available in GPI pipeline \citep{maire10, perrin14} v. 1.3, with notable customizations. First, microphonics noise was minimized by eye through tuning parameters in the ``destripe science image" primitive. Second, the ``subtract mean stellar polarization" primitive was added to the standard polarization recipe to remove instrumental polarization (estimated in the region $1<$ r $< 10$ pixels, which shows no significant disk emission). Finally, a custom pipeline fix was implemented to allow image alignment in unblocked mode. GPI image alignment typically relies on well-calibrated satellite spots injected by the GPI apodizers which are not present in unblocked mode, so a Gaussian stellar centroid was used instead. Though this alignment method is imperfect due to stellar saturation, it worked well in this case. The resulting images were transformed to ``radial" Stokes parameters (Schmid et al. 2006). Images used here represent the Q$_R$ component, which holds all linear polarized flux oriented either parallel (negative) or perpendicular (positive) to the line connecting that pixel to the central star. Positive flux in these images therefore represents singly-scattered photons from the circumstellar disk.   

HIP~56071 was observed as a flux standard, with the same procedures, except without closing the AO loop, thereby avoiding saturation. This allowed us to derive conversion factors between ADU sec$^{-1}$ pixel$^{-1}$ and mJy asec$^{-2}$. Since the K1-band magnitude in HIP~56071 was unavailable, we translated the 2MASS K$_{\rm s}$-band magnitude into this band by relating the stellar flux to the Vega flux assuming box passbands in two bands (the GPI K1-band with 1.9-2.19 $\mu$m and the 2MASS K$_{\rm s}$-band with 1.989-2.316~$\mu$m) and blackbody radiation with a T$_{\rm eff}$ of 9200~K for HIP~56071 (A1V) and 9700~K for Vega (A0V). The color correction of K$_{\rm s \ 2MASS} - $K1$_{\rm GPI}$ was $-$0.002; thus, we derived HIP~56071's K1$_{\rm GPI}$ magnitude is 7.860. We derived conversion factors of 1~ADU sec$^{-1}$ pixel$^{-1}$ are 0.846 and 0.853 mJy asec$^{-2}$ in the J and K1-bands, respectively. Note that Y-band flux was not converted to mJy asec$^{-2}$ due to no literature value for the Y-band. 

We find the projected separation between the image center and the companion to be $1\farcs05$, in agreement with \citet{chen06}. The region within $\sim$4 pixels (corresponding to $\sim$0$\farcs$06) from the central star is saturated.  We conservatively estimate the features outside 6-7 pixels from the center to be real while the area interior to this radius is washed out by speckle residuals. We detect the outer disk and spirals, as in \citet{wagner15}. The total PI within $0 \farcs 1 < r < 1 \farcs 0$ (the inner radius corresponds to 7 pixels) is 13~mJy and 22~mJy for J- and K1-bands, respectively.

\subsection{Archival Total Intensity Imagery and Assembly of the IR SED}
\citet{wagner15} also observed HD 100453 with VLT/SPHERE on 2015 April 10. The observations were carried out in IRDIFS extended mode\footnote{Very Large Telescope SPHERE User Manual} \citep{gpimanual} using IRDIS to take dual-band TI images in K1- and K2-bands and simultaneously using the IFS to obtain low-resolution spectra from Y- to H-bands. Data reduction is described in \citet{wagner15}. Photometry used in constructing HD 100453's SED includes all sources mentioned in \citet{khalafinejad16} as well as \textit{Herschel} PACS data at 70, 100, and 160 $\mu$m \citep{pascual16}, and 2MASS at J-, H-, and Ks-bands \citep{cutrie03}.

\section{RESULTS}

\subsection{SPHERE and GPI Imagery}
\begin{figure}[ht]
\centering
\includegraphics[scale=0.5]{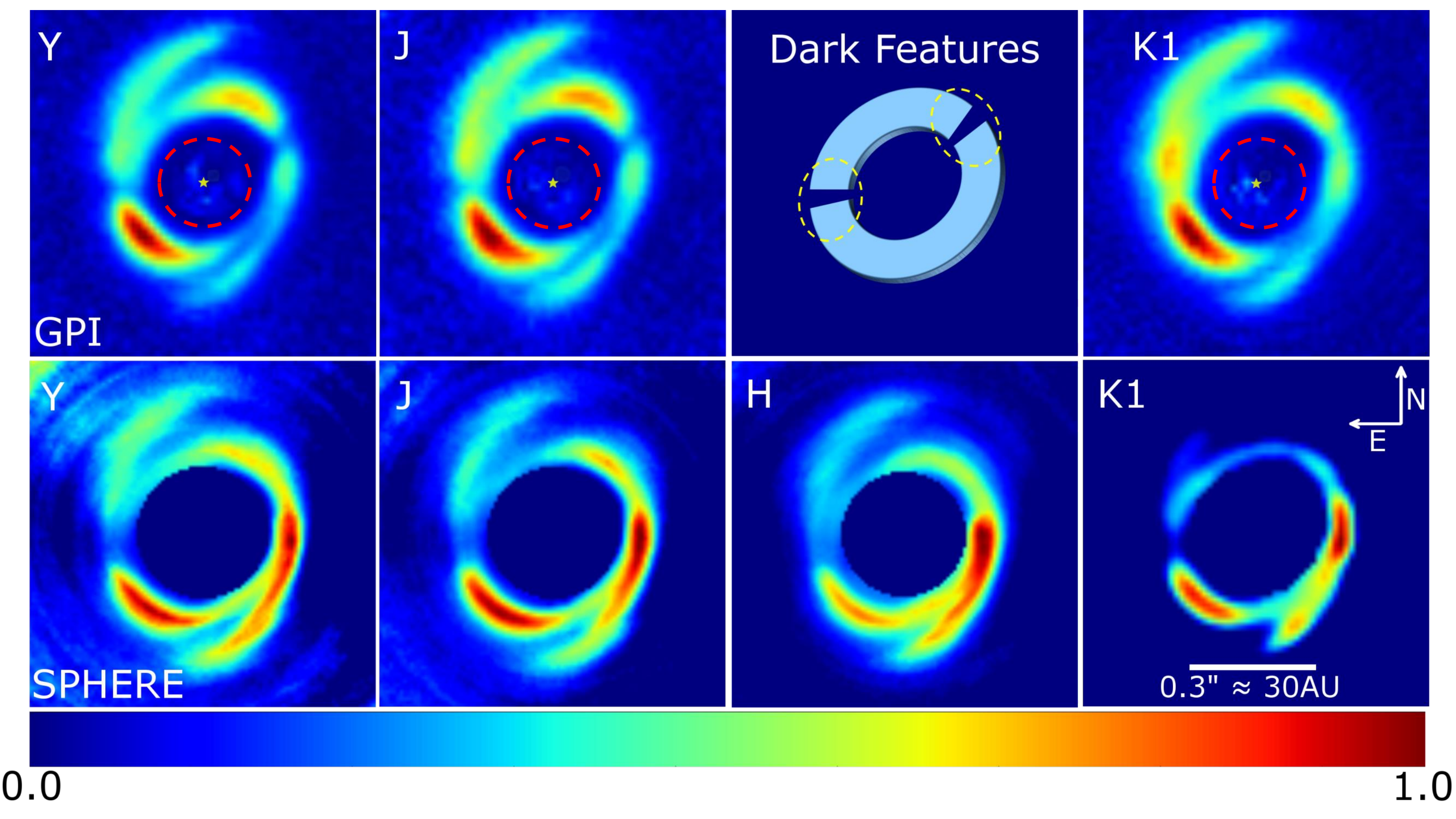}
\caption{The r$^2$-scaled GPI polarized intensity images in Y, J, and K1 bands, plus a schematic of the disk marking the locations of the dark features (top row), and r$^2$-scaled SPHERE total intensity images of HD 100453 in Y, J, H, and K1 bands (bottom row). The images have a  field of view of $0\farcs8\times0\farcs8$ and are normalized by the maximum intensity in each band. The central star is represented by a yellow star in GPI images and the inner working angle is represented by a red dashed line.}
\label{fig:r^2 SPHERE and GPI}
\end{figure}

A radial gap can clearly be seen from $9 \pm 2$ au, to $18 \pm 2$ au in the GPI imagery (Figure~\ref{fig:r^2 SPHERE and GPI}). Inside 9 au the GPI image is saturated, making the inner edge of the gap inaccessible in the image. The outer radius agrees with the \citet{wagner15} estimate of $\sim$19 au using $d = 103 \pm 3$ pc \citep{gaia16}, and the \citet{khalafinejad16} estimate of $20 \pm 2$ au from SED modeling. We measure a pericenter offset upper limit of $1 \pm 1$ pixels (0$\farcs014 \pm 0\farcs014$) (Figure~\ref{fig:r^2 SPHERE and GPI}). 

The outer disk is brightest on the southern side in both PI and TI imagery implying this is the near side of the disk. However, the northern spiral arm is brighter in PI images while the southern arm is brighter in TI images, indicating possible differences in grain properties. Additionally there is a dropoff in the spiral arm intensity at longer wavelengths in SPHERE TI images which suggests they made of smaller grains than the outer disk. After absolute flux calibration of the TI images using archival 2MASS photometry \citep{cutrie03} we measure the total TI flux between $0 \farcs 1 < r < 1 \farcs 0$ is 49 mJy and 70 mJy in J- and H-bands respectively. Using this and the 2MASS flux we calculate a fractional luminosity for the outer disk \(\frac{fdisk}{ftotal}\) of $0.018 \pm 0.001$ for J-band and $0.025 \pm 0.002$ for H-band which may suggest that the outer disk is reddish and therefore comprised of large, compact grains \citep{mulders13}.  By contrast we calculate a fractional luminosity for the spiral arms of $0.0040 \pm 0.001$ for J-band and $0.0045 \pm 0.001$ for H-band which, though slightly red, is more blue than the rest of the outer disk indicating they may be comprised of smaller grains \citep{mulders13}.

Two distinct azimuthally-localized dark features are seen at the same position angle (PA) in both TI and PI images, and are therefore not artifacts of PI imagery. They also have a similar outer disk to feature contrast in all available bands (Figure~\ref{fig:r^2 SPHERE and GPI}).

\subsection{Determining the Likely Cause of the Dark Features}

We discuss three possible physical scenarios for the origin of the dark features: a physical gap due the dynamical clearing of a large body, grain growth and settling, or shadows cast from an inner disk component.

The SPHERE Y-band PSF for HD 100453A has a FWHM of ~5 pixels (~3.8 au). This suggests the dark features are resolved and have an azimuthal extent of $14 \pm 2$ pixels ($10 \pm 1.5$ au) based on the FWHM of the eastern feature in this image. If the dark feature is due to the clearing of a single body's Hill sphere \citep{hamiltonandburns92}, it would correspond to an M-type star with a mass of at least 0.4 M$_{\astrosun}$. The M-type companion, HD 100453B, is visible in SPHERE, GPI, and Chandra \citep{collins09} imagery. An M-type star with a mass of 0.4 M$_{\astrosun}$ therefore would certainly be visible in the location of the dark features if it existed. Thus, we reject the hypothesis of local clearing to explain the dark features. 

Grain growth and settling has been suggested as a source of dark regions at NIR wavelengths \citep{dullemondanddominik04a, dullemondanddominik04b, birnstiel12}, and should also produce bright rings in the sub-mm. This would require resolution of at least $0\farcs03$, which is reachable by interferometric telescopes such as ALMA. The differential rotation of the disk would cause the dark features to deform over time however, which suggests that grain growth and settling is not the cause of the features. 

An inner disk would cast two shadows which have a large radial extent and have similar outer disk to shadow contrast over the wavelength range in which it is optically thick \citep{stolker16}, similar to what is seen in the GPI and SPHERE images. We will show that shadows cast by an inner disk with a suitable inclination is fully capable of producing such dark features. In the following sections we present a model that is capable of generating the dark features as well as reproducing the observed SED.

\subsection{Development of Preliminary Model}

\subsubsection{Literature Values for Outer Disk Inclination}

\citet{dong16} adopted $i\sim$ 5$\degr$, where $i$ is the outer disk inclination from face on with respect to the observer, to reproduce the spiral arm morphology in hydrostatic modeling of the HD 100453 system. If we accept the \citet{dong16} assumption of a completely coplanar disk, we can also assume the equatorial plane of the star is coplanar because, according to \citet{greaves14}, most stars rotate in the same plane as their disks. Using the 5$\degr$ inclination for the outer disk and a $v \, sin \, i$ of $48 \pm 2$ km s\textsuperscript{-1} for HD 100453A \citep{guimaraes06} we find the equatorial velocity of HD 100453A would be $\sim$550 km s\textsuperscript{-1}, which is $150$ km s\textsuperscript{-1} above the break-up velocity of the star \citep{slettebak66}. Because HD 100453A is still intact, a 5$\degr$ inclination for the star is non-physical which implies, by extension, that the inclination of the outer disk also cannot be $5\degr$. Moreover in GPI images we can clearly see the major and minor axes of the disk which would not be possible given a nearly face on disk as in TW Hydrae \citep{andrews16}. If the spiral arms are close to face on as \citet{dong16} suggest, they are not coplanar with the outer disk.

\citet{wagner15} measured an outer disk inclination of $i\sim$ 34$\degr$ from face on through fitting an ellipse to the peak intensity along the center of outer disk and assuming circular geometry. However, because the outer disk has finite thickness, this method will overestimate the outer disk's inclination. We will review this inclination in Section 3.3.3.

\subsubsection{Initial Modeling Parameters}

\begin{figure}[ht]
\centering
\includegraphics[scale=1]{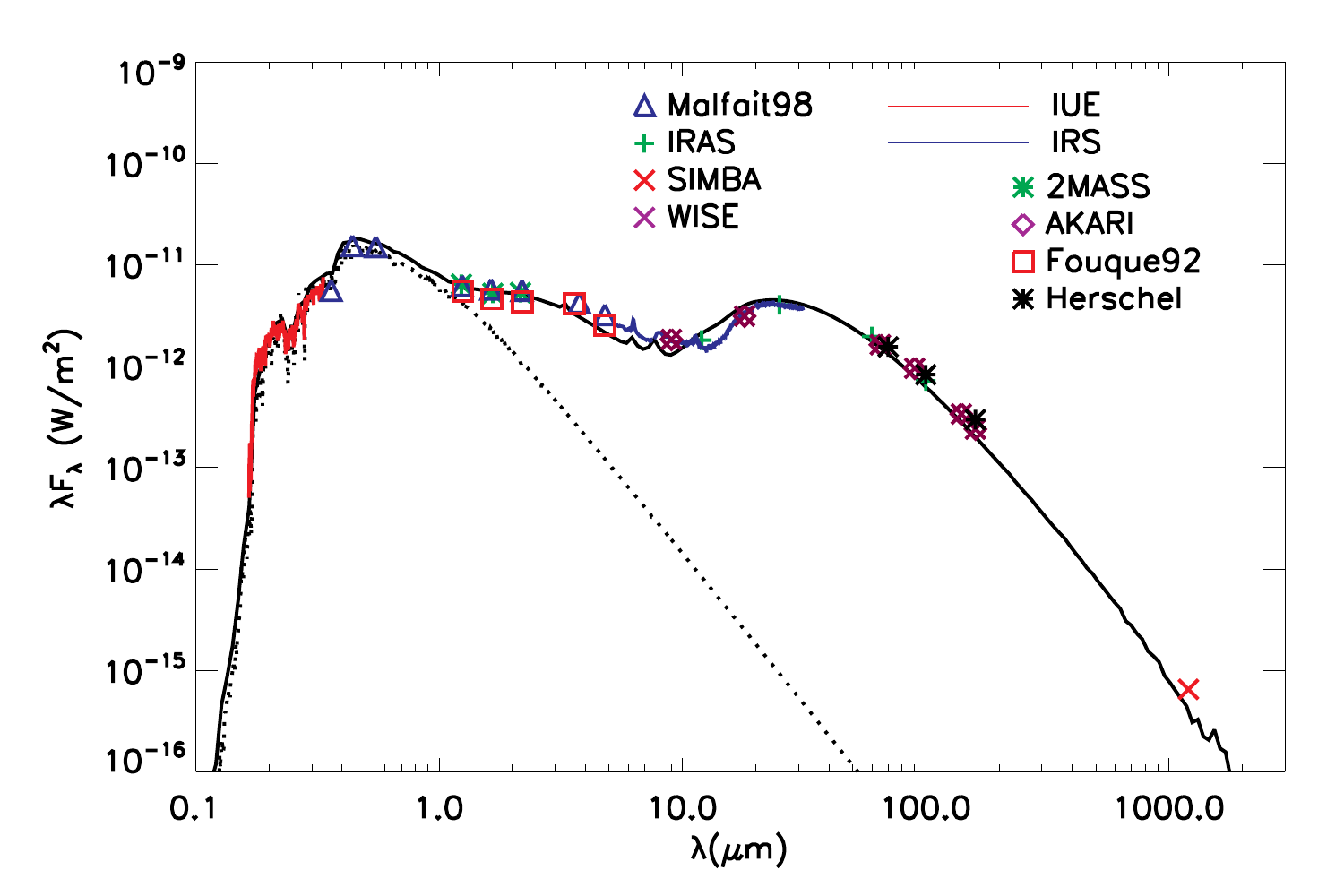}
\caption{Best model fit (solid line) of the SED of HD 100453 after initial modeling with the stellar atmosphere (dotted line) also visible.}
\label{fig:SED}
\end{figure}

In order to reduce the degeneracies in our model, literature values, SPHERE and GPI imagery, and the SED (Figure~\ref{fig:SED}) are used in combination to determine initial modeling parameters. For the star we used a PHOENIX stellar atmosphere model \citep{brott05} with T = 7400 K (appropriate for an A9 star) and a distance of 103 pc \citep{gaia16}. HOCHUNK3D uses the \citet{lucy99} method for calculating temperature in our model. Box filters were created in Y-K2 to allow for direct comparison of the model to data.

Through SED fitting \citet{khalafinejad16} suggested that the the inner disk extends from 0.25 to 1.7 au. Confirmation of the inner edge radius is provided by H-band interferometry with VLTI/PIONIER \citep{lazareff16} who find an upper limit of 0.27 au. The inner disk extends to at least $0.9 \pm 0.1$ au from the N-band half light radius of VLTI/MIDI \citep{menu15}, updated using $d = 103 \pm 3$ pc \citep{gaia16}. The long wavelength slope obtained from the 1.2 mm SIMBA point\citep{meeus02b}, the reddish color of the outer disk (Section 3.1), and the lack of a strong silicate peak in the SED \citep{meeus02a}, imply that the inner and outer disks are comprised of large, compact dust grains \citep{mulders13}. 

To match these dust properties we used Model 2 from \citet{wood02} for the settled disk grain opacities. This contains a mixture of amorphous carbon and astronomical silicates, a radial power-law exponent (a) of 3.5 for both grain compositions, no exponential cutoff to the grain size distribution, a a maximum particle size $\leq$ 1 mm, and a minimum particle size of 0.01 $\mu m$. For the less-settled grain opacities we used Model 1 from \citet{wood02}.  Similarly this contains a mixture of amorphous carbon and astronomical silicates, a power law size distribution with a = 3.5 and 3.0, respectively, plus an exponential cutoff with a turnover at 50 $\mu$m, a maximum particle size $\leq$ 1 mm, and a minimum particle size of 0.01 $\mu m$. 

\subsubsection{Inclination of the Outer Disk}

Because reproducing the spiral arms is not necessary for studying the cause of the dark features, they are not included in our models. We calculated models with $5\degr \leq i \leq 45\degr$ in 5$\degr$ increments and compared the resulting model images to observed images and find a best fit of $25 \pm 10\degr$. From this model we also find a PA for the outer disk's major axis of $140 \pm 10\degr$ east of north. 

\subsection{Adopted Model Parameters}

After initial modeling of the outer disk, we constructed a more complete disk model to fit the SED (Figure~\ref{fig:SED}). We find that our model closely matches the observed SED including the NIR region, which was ignored by \citet{benisty17}, to be discussed later. It consists of an inner disk ($\sim0.13 - 1 \pm 0.5$ au) which reproduces the IR excess of the SED and is in agreement with VLTI/MIDI measurements of $0.9 \pm 0.1$ au. This is followed by a depleted region to 18 au, and finally an outer disk ($18-39$ au). The inner disk has a vertical inner-edge thickness of 0.11 $\pm$ 0.05 au as defined by $z = Cr^b$ where z is the density scale height (thickness) of the disk, C is a constant, r is the radial distance from the star, and b is the flaring exponent \citep{whitney03a}. To match mid-far IR emission we find b = 1.28 $\pm$ 0.02 which falls within the canonical values of 1.25 - 1.3 for irradiated disks \citep{chiang97, hartmann98}. Changes in the settled disk have little effect on the SED and were not changed appreciably in our modeling. Our final parameters for the settled disk were z = 0.0004 au and b = 1.25. Both the vertical density and the surface density profile are as described in \citep{whitney13} for a disk in hydrostatic equilibrium. In particular for the less settled disk and settled disks we chose density exponents $\alpha = 2.30$ and $\alpha = 2.25$ respectively.

\subsection{Difference in Inclination for Inner Disk}

\subsubsection{Definition of $\Delta i$}
\begin{figure}[ht]
\centering
\includegraphics[scale=0.6]{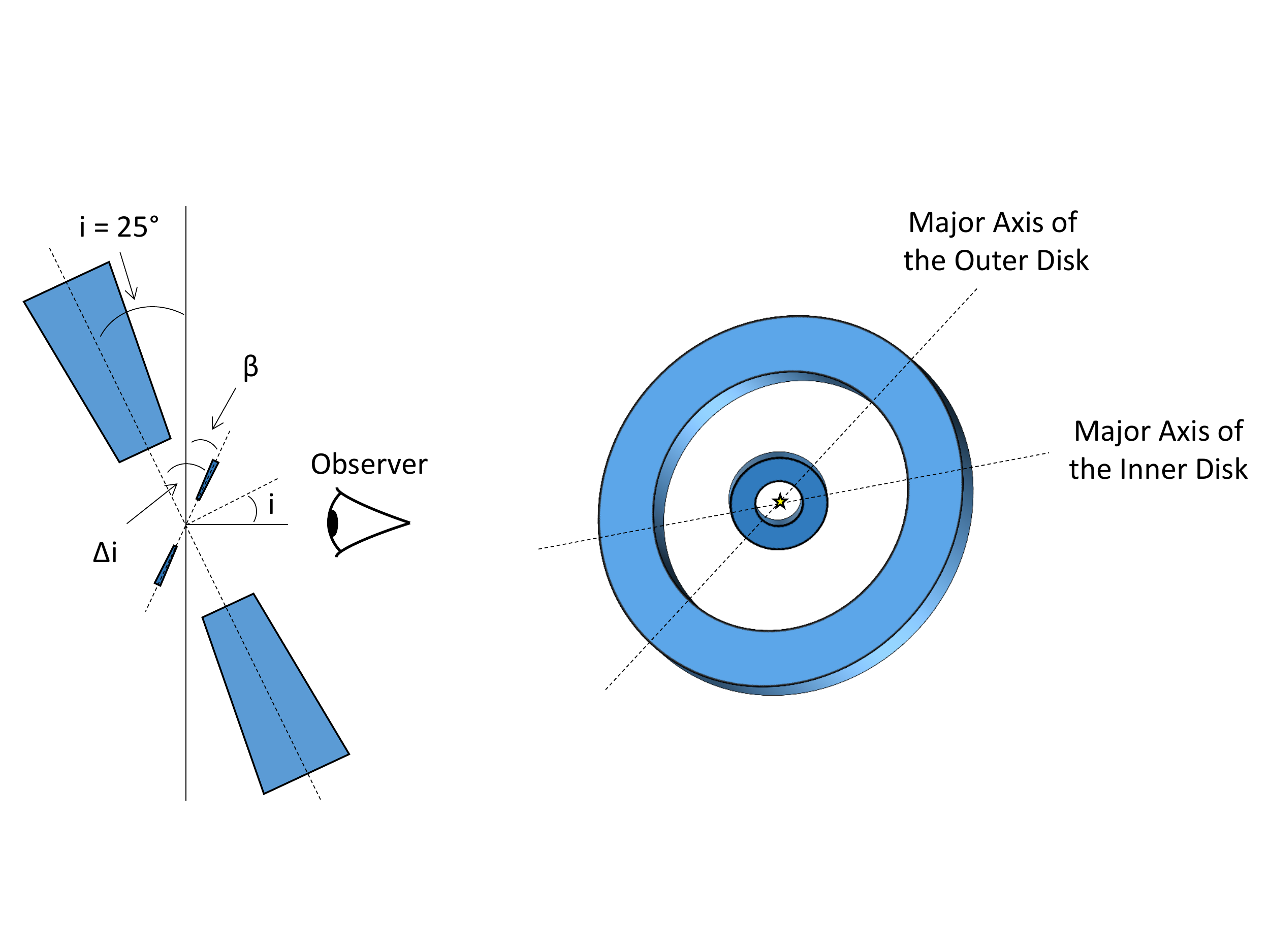}
\caption{LEFT: A cross section of the disk structure with $\Delta i = i - \beta$ where $i$ is the positive inclination of the outer disk from face on with respect to the observer while $\beta$ is the positive or negative inclination of the inner disk from face on. RIGHT: A schematic view of the overall disk structure of HD 100453 in the observed frame, as deduced from our best-fit model (Section 3). The major axes of the inner and outer disks are marked.}
\label{fig:Schematic of Disks}
\end{figure}

In this section we test the validity of a difference in inclination between the inner and outer disk as the cause of the dark features. We define this difference in inclination as $\Delta i = i - \beta$ where $i$ is the inclination of the outer disk from face on and $\beta$ is the inclination of the inner disk from face on. While we define $i$ as positive, $\beta$ can be positive or negative depending on the inner disk's direction of tilt (Figure~\ref{fig:Schematic of Disks}). 

\subsubsection{Azimuthal Separation of Dark Features and Comparison to Model}
\begin{figure}[ht]
\centering
\includegraphics[scale=0.1]{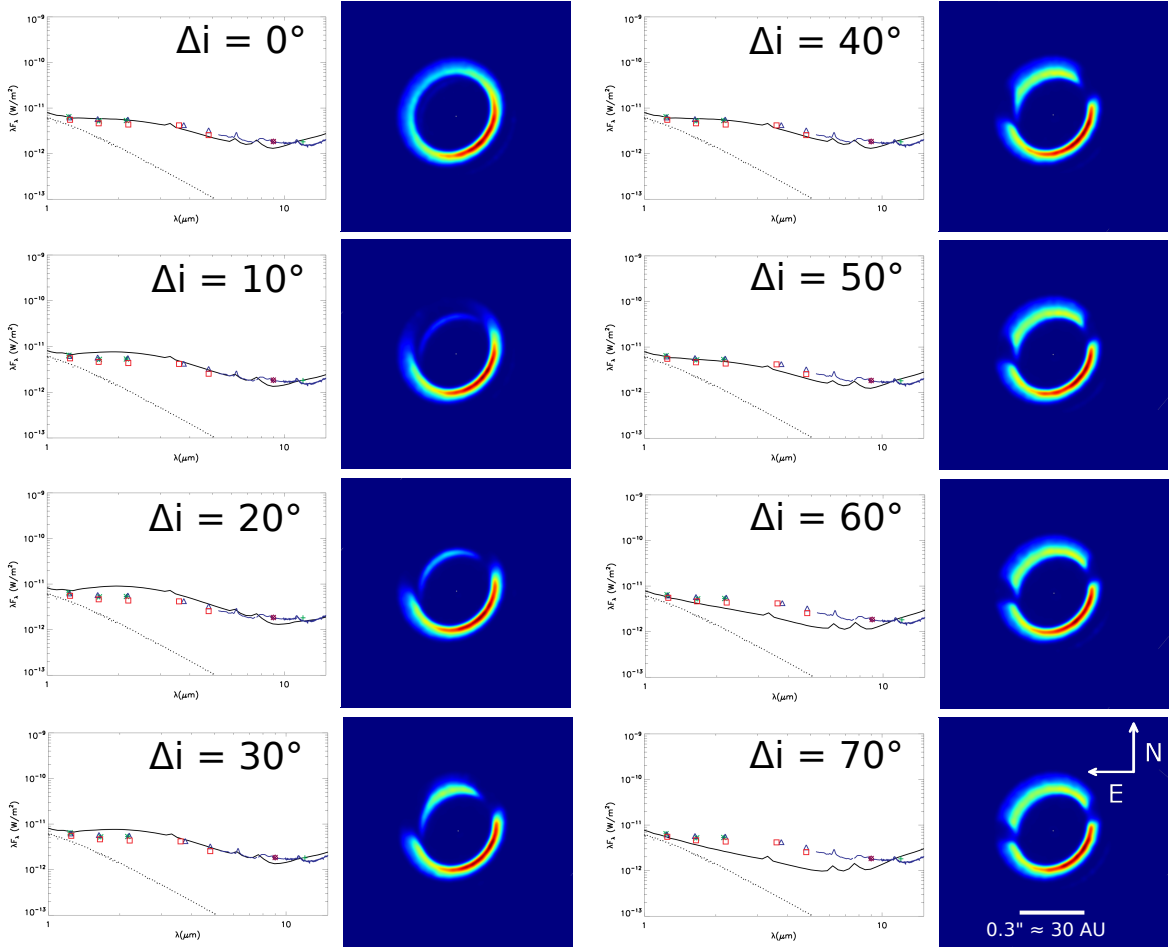}
\caption{SEDs between 1 and 10 $\mu$m coupled with r$^2$-scaled, total intensity model images at J-band, convolved with the PSF of SPHERE total intensity imagery, of various $\Delta i$'s. Here the $\Delta i = 0$ does not produce dark features and therefore HD 100453 cannot have two coplanar disks. We find a best fit of $\Delta i = 45 \pm 10\degr$.}
\label{fig:Delta i model and sed}
\end{figure}

Model images were produced with the adopted $i = 25\degr$ while $\Delta i$ is varied in 10$\degr$ increments from 0$\degr$ to 70$\degr$ (Figure~\ref{fig:Delta i model and sed}). These are convolved with the PSF of SPHERE J-band images and therefore produce clearer shadows than longer wavelength bands. No azimuthally-localized dark features were observed in model images with $\Delta i = 0\degr$ despite a good fit to the SED, excluding a coplanar disk system. At $\Delta i = 20\degr$ the inner disk shadows the northern section of the image and predicts too much flux between 1-10$\mu$m because of its nearly face-on orientation with respect to the observer (Figure~\ref{fig:Delta i model and sed}). Moving to $\Delta  i =  70\degr$ the shadows narrow and predicts too little flux between 1-10$\mu$m because the inner disk is nearly edge-on to the observer. We find $\Delta i = 45 \pm 10\degr$ matches the general appearance of the imagery, to be quantified below.

\begin{figure}[ht]
\centering
\includegraphics[scale=0.25]{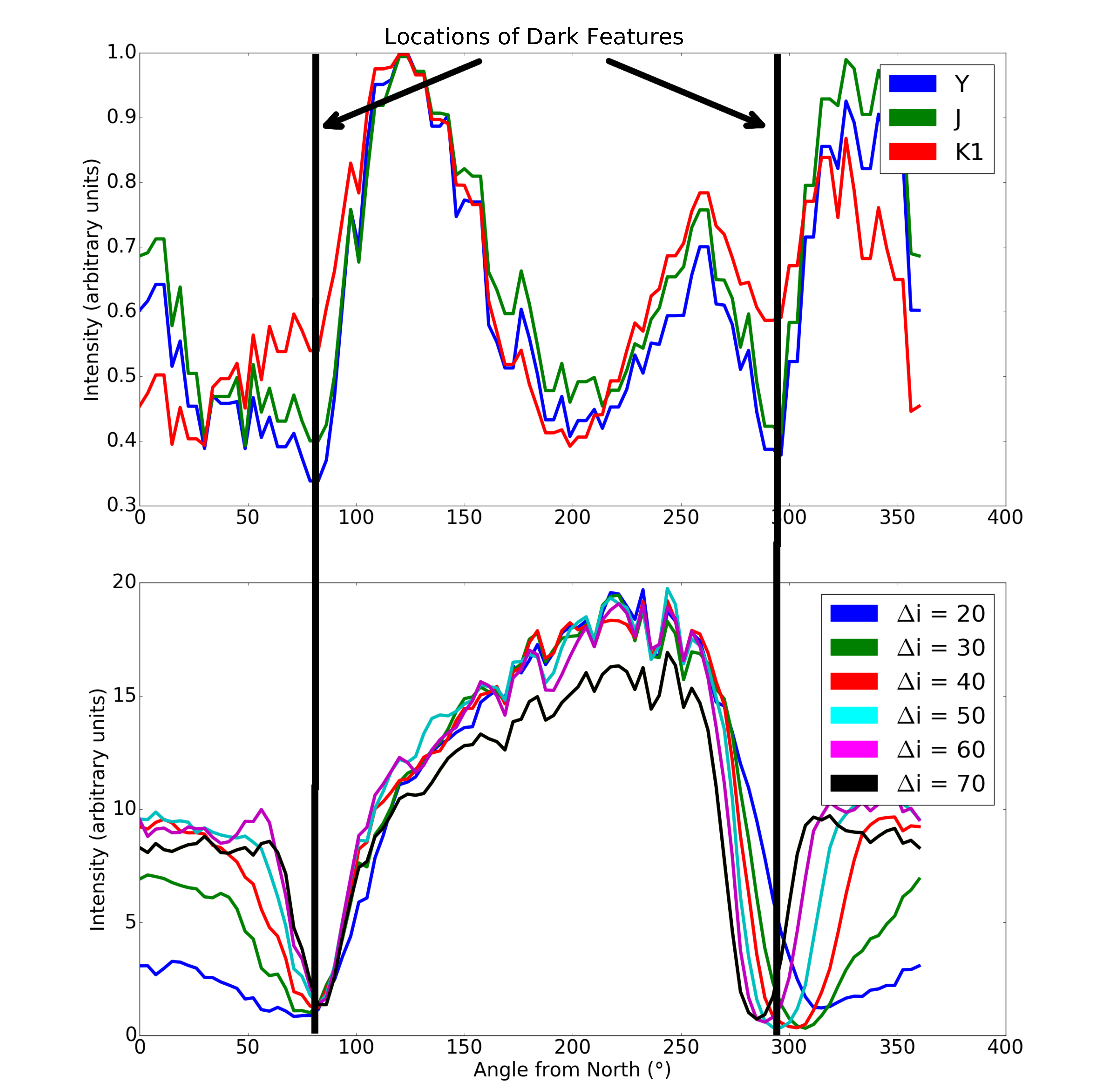}
\caption{The colored lines represent azimuthal intensity profiles of GPI polarized intensity (top) and total intensity, J-band model (bottom) images where 0$\degr$ represents due North and we trace counterclockwise along the outer disk. GPI traces are normalized by the maximum intensity in each band. Model images are convolved with the PSF of SPHERE imagery and unscaled. The vertical black lines mark the position of the dark features.}
\label{fig:Separation of Dark Features}
\end{figure}

Azimuthal intensity profiles of the GPI images and model imagery were generated for the outer disk (Figure~\ref{fig:Separation of Dark Features}) and measure an azimuthal separation of $140 \pm 10\degr$ for the dark features which are in the same locations in each band. The additional local minimum in intensity profile plots is due to polarization effects along the minor axis of the outer disk. We find that as $\Delta  i$ is increased, the shadow separation also increases and produces a best fit to the dark features at $\Delta  i = 45 \pm 10\degr$ which corresponds to $\beta = -20 \pm 10\degr$. This deduced inner disk inclination angle is significantly different from the one found in \citet{lazareff16}, $\beta = -48\degr$, using H-band interferometry with VLTI/PIONIER. The main source of uncertainty in our measurement comes from azimuthal extent of the shadowed region. To match the dark features' azimuthal location we rotate the inner disk by -40$\degr$ with respect to the outer disk which places the major axis of the inner disk at $100 \pm 10\degr$ east of north. This differs from the PA for the major axis found in \citet{lazareff16}, 81$\degr$, and will be discussed later. 

\begin{figure}[ht]
\centering
\includegraphics[scale=0.5]{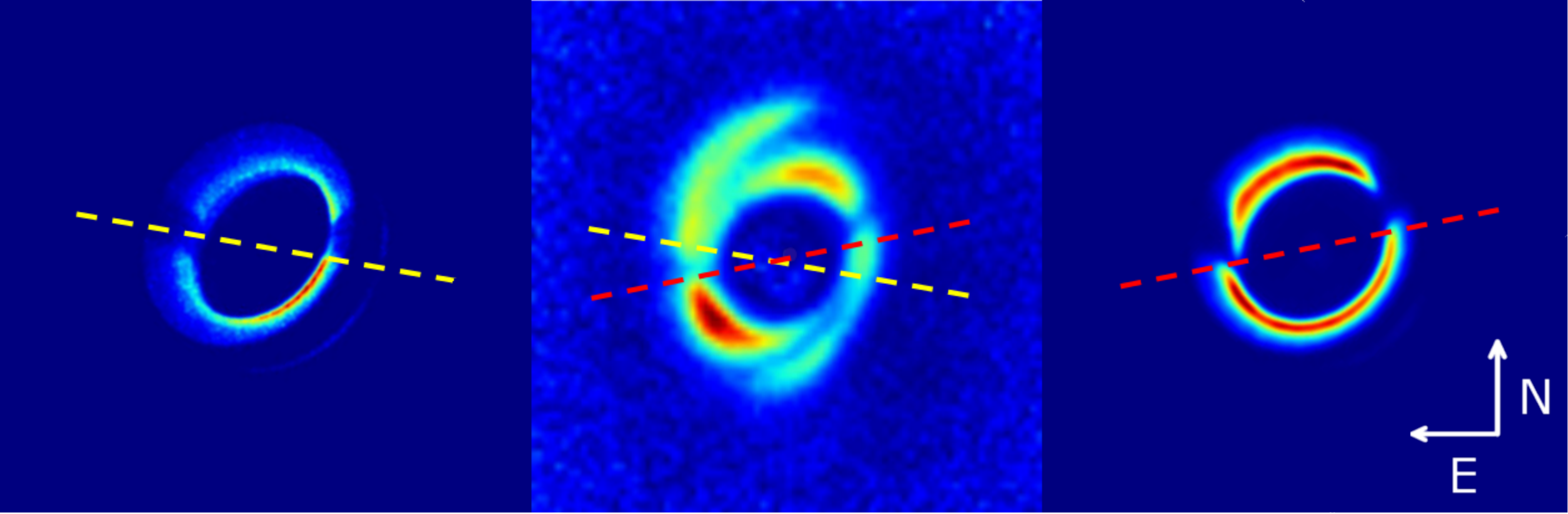}
\caption{LEFT: The r$^2$-scaled, total intensity \citet{benisty17} model recreation at J-band, convolved with the PSF of SPHERE total intensity imagery. CENTER: Observational GPI polarized intensity r$^2$-scaled J-band image. RIGHT: r$^2$-scaled, total intensity model image at J-band of our best-fit model outlined in Section 3.5, convolved with the J-band PSF of SPHERE total intensity image. Here the yellow dotted line shows the major axis PA of the inner disk in the \citet{benisty17} model recreation and the red dotted line represents the major axis PA of the inner disk in our model. It is apparent from the figure that the \citet{benisty17} model recreation that neither the major axis location nor the shadow locations themselves match the locations of the shadows in the data. In addition simple visual examination of this model reveals that both the azimuthal separation of the shadows and the ellipticity are too large to match the observational data.}
\label{fig:Benisty}
\end{figure}

A recent study by \citet{benisty17} also proposed a misaligned inner disk as the cause of the outer disk shadowing in HD 100453. In their paper they generate a model to test this hypothesis using measurements from \citet{lazareff16}. In order to test the validity of their model we generated our own using their value for the outer disk inclination, $i = 38\degr$, and their quoted $\Delta i$ of  $72\degr$. This model however does not produce shadows in the same locations as in the data (Figure~\ref{fig:Benisty}), similar to what was found earlier in this section. 

\begin{figure}[ht]
\centering
\includegraphics[scale=0.28]{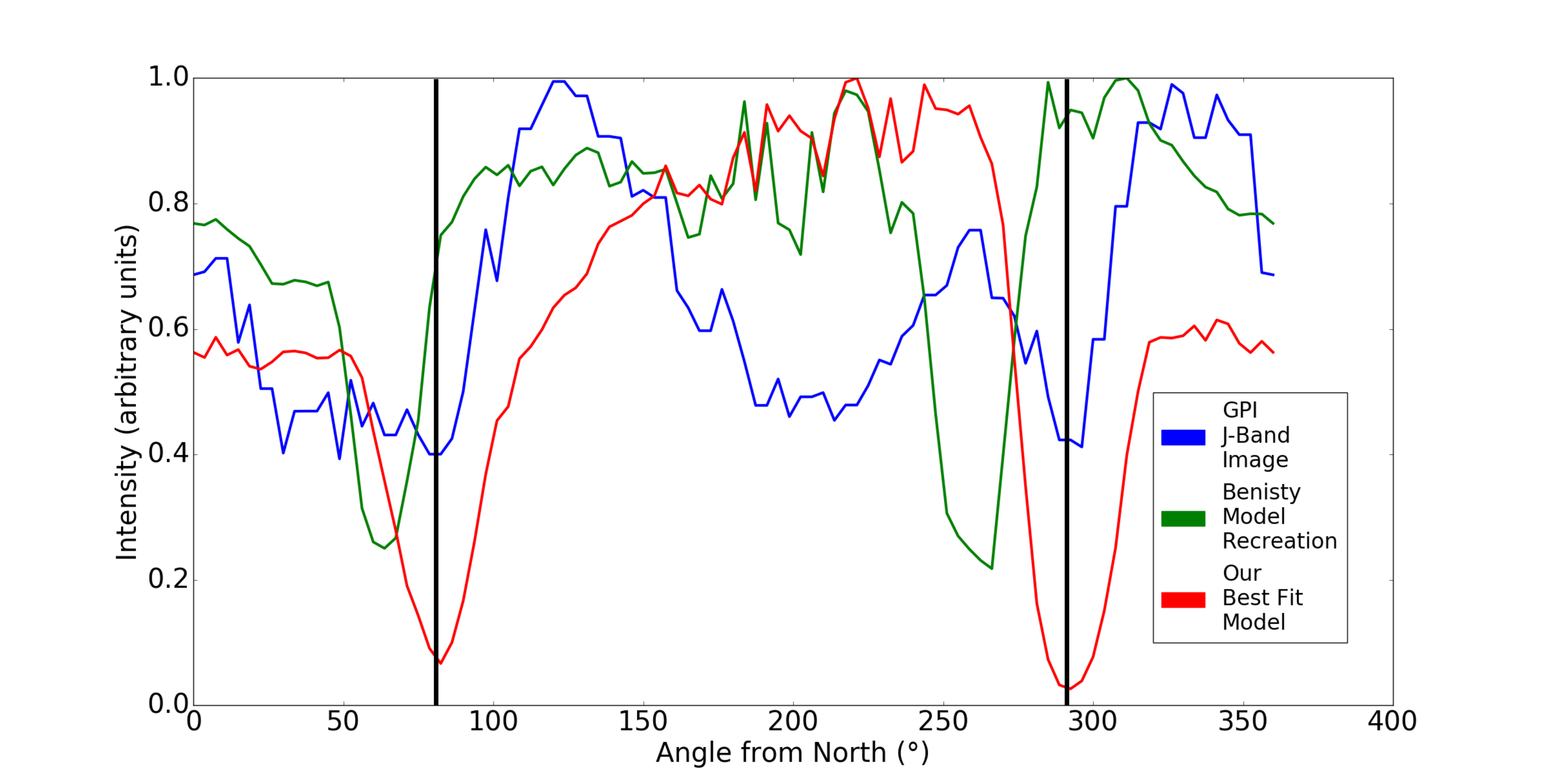}
\caption{The colored lines represent azimuthal profiles of GPI polarized intensity, our best fit J-band model, and our recreation of the \citet{benisty17} model. The vertical black lines represent the location of the shadows in the GPI image. In our model recreation using $\Delta i = 72\degr$ and i = $38\degr$ neither the location of the shadows nor their separation agree with the GPI imagery.}
\label{fig:bensitytrace}
\end{figure}

In our modeling we find that the inclination of the outer disk had little effect on the separation of the shadows on it's own unless the near side of the outer disk is reversed. We also find that the separation of the shadows is less than $180\degr$ on the side of the outer disk which corresponds to the near side of the inner disk. Essentially the inner disk major axis divides the outer disk into two semicircles. The $180\degr$ constraint implies that the centers of the shadows cannot exist in separate semicircles and this strongly constrains the orientation of the inner disk. In Figure~\ref{fig:Benisty} we see that the quoted orientation of the inner disk major axis places the shadows in separate semicircles in the GPI image which cannot occur if the inner disk is the source of the shadows. In addition the gap has a visibly more elliptical structure in the \citet{benisty17} model than in the data which is likely due to the thickness of the outer disk as discussed in Section 3.4 as well as the larger inclination of the outer disk. It is important to note that this $\Delta i$ corresponds to $\beta = -34\degr$ instead of the inner disk inclination proposed separately in \citet{benisty17} of $\beta = -48\degr$. Additionally in our modeling we find that the NIR region of the SED is fairly sensitive to the inclination of the inner disk (Figure~\ref{fig:Delta i model and sed}). Because \citet{benisty17} did not fit the NIR excess of the SED, the \citet{benisty17} model is not as tightly constrained as our simultaneous image and SED fitting. We found the inclination of the inner disk proposed \citet{benisty17} did not produce a good match to the 1-10 micron SED nor the constraints on the shadows that that we found (Figure~\ref{fig:bensitytrace}). Differences in the inner disk orientation which produce shadows along an axis other than the major axis may be possible, though we did not observe this in our modeling.

\subsection{Effects of Inner Disk Thickness and Outer Disk Flaring on Shadows}

\begin{figure}[ht]
\centering
\includegraphics[scale=0.25]{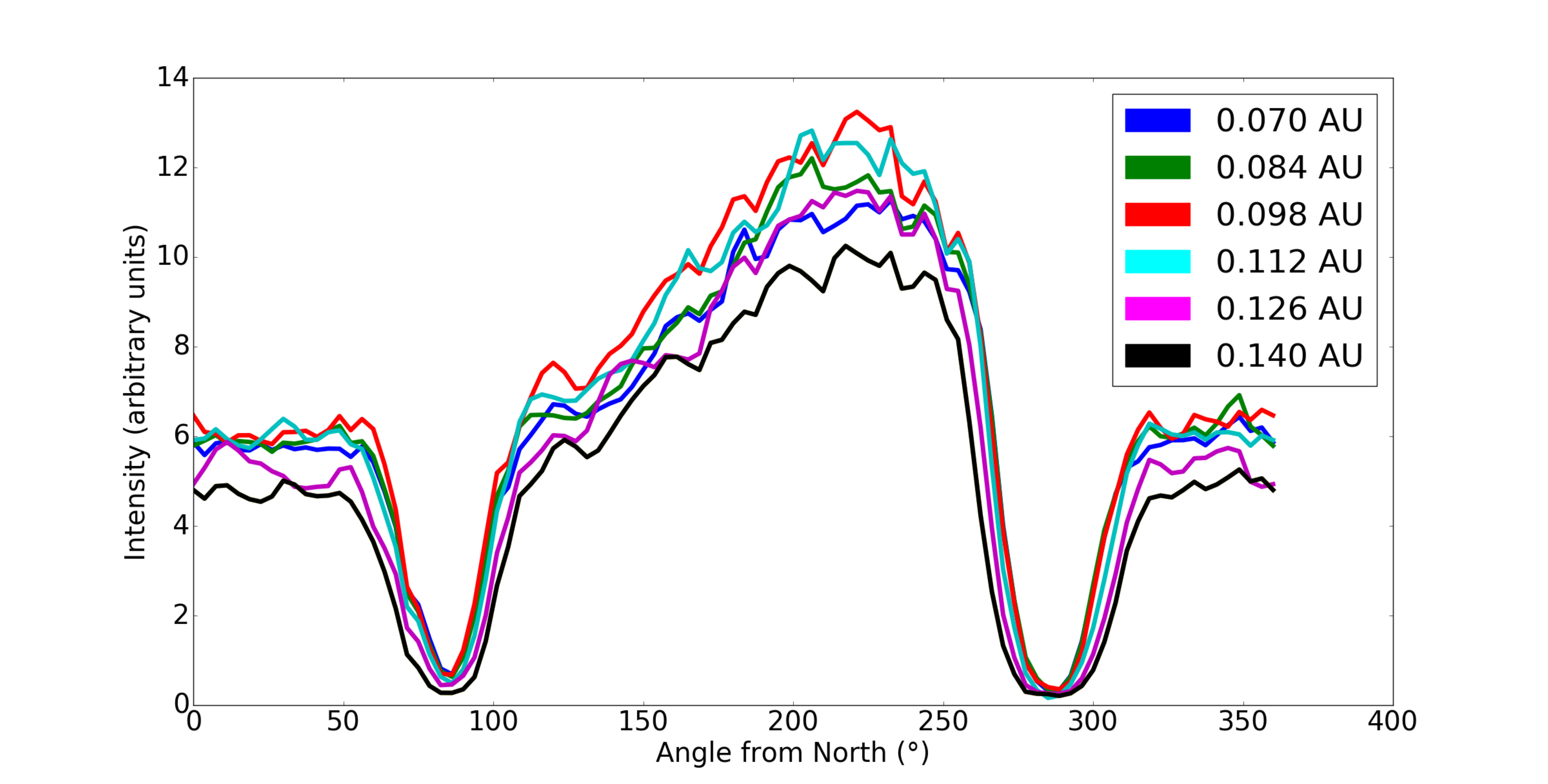}
\caption{Graph showing the effects of inner disk thickness on the width of the dark features. The colored lines represent intensity traces derived from unscaled J-band total intensity model images convolved with the PSF of SPHERE total intensity imagery. 0$\degr$ represents due North and we trace counterclockwise along the outer disk. We find that as the thickness increases the width of the dips associated with the shadows also increases while their intensity decreases.}
\label{fig:Thickness vs Width}
\end{figure}

To examine how inner disk thickness affects the morphology of the shadows, we generated model images at inner-edge inner disk thicknesses from 0.07 - 0.14 au in increments of 0.014 au without fitting the SED. We find that as the thickness increases, the width of the shadows increases with no change in the location of the shadows. (Figure~\ref{fig:Thickness vs Width}). The differences in width become small at smaller thicknesses however, suggesting that shadows can act as an upper constraint of inner disk thickness. 

\begin{figure}[ht]
\centering
\includegraphics[scale=0.65]{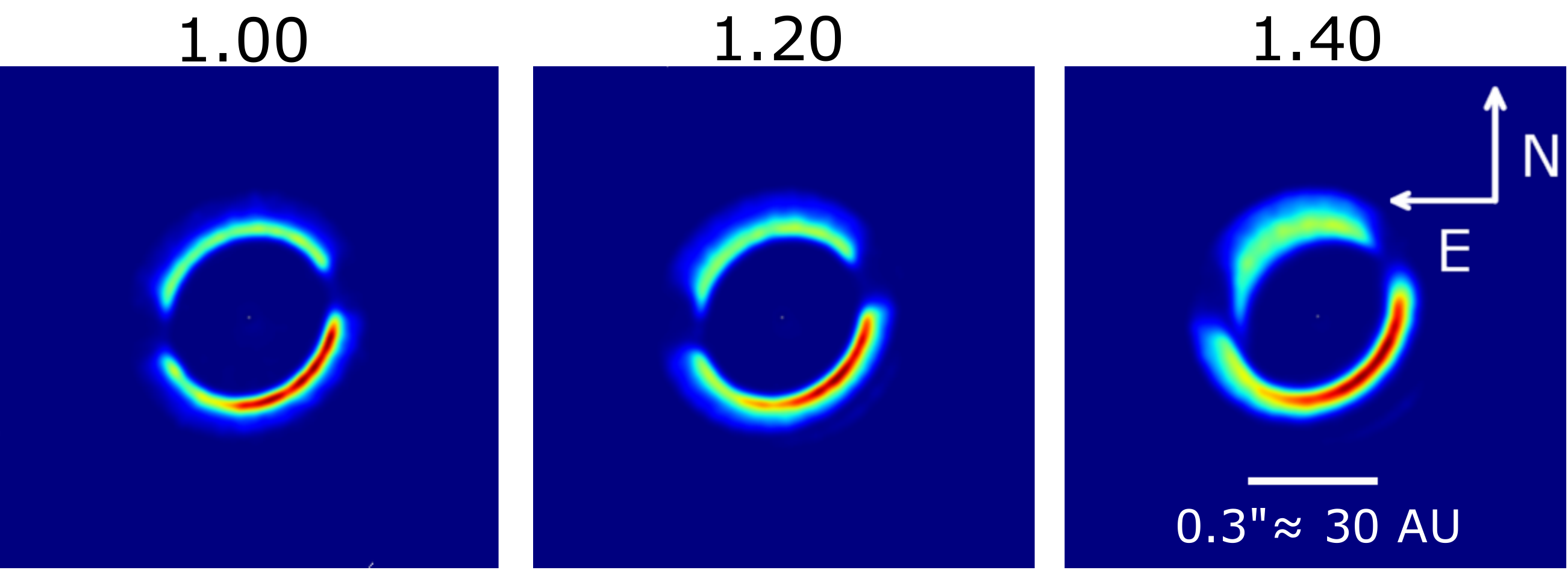}
\caption{Figures showing total intensity model generated images at J-band with flaring exponents (Section 3.4) of 1.00, 1.20, and 1.40. Model images are convolved with the PSF of SPHERE imagery and unscaled. The SW side is the near side of the outer disk and it is clear that the near edge is thinner in the flared disks than the far edge.}
\label{fig:Flaring}
\end{figure}

Model images were also generated at flaring exponents (Section 3.4) of 1.00, 1.20, and 1.40 (Figure~\ref{fig:Flaring}) in order to examine the effect of outer disk structure on shadow morphology, also without fitting the SED. We find that the width of the shadows' inner edge decreases with increasing b, while the outer edge remains largely unaffected. In addition we observe that the flaring of the disk causes the outer disk to appear thicker on the far side as seen in Figure~\ref{fig:Flaring} where the SW side is the near side of the outer disk. This effect is most prevalent in the rightmost panel where the SW side of the disk is approximately half the thickness of the NE side. In the case of HD 100453, the SW side of the outer disk is narrower in GPI images (Figure~\ref{fig:r^2 SPHERE and GPI}) indicating that this is the near side. The ratio of thicknesses between the near and far side of the disk, coupled with the inclination of the outer disk, could allow us to quantitatively describe the degree of flaring of the outer disk. When taken in conjunction with the inner disk thickness, this should allow us to strongly constrain the disk's physical structure. 

\subsection{Using Shadows to Determine Near Side of Inner and Outer Disk}

\begin{figure}[ht]
\centering
\includegraphics[scale=0.7]{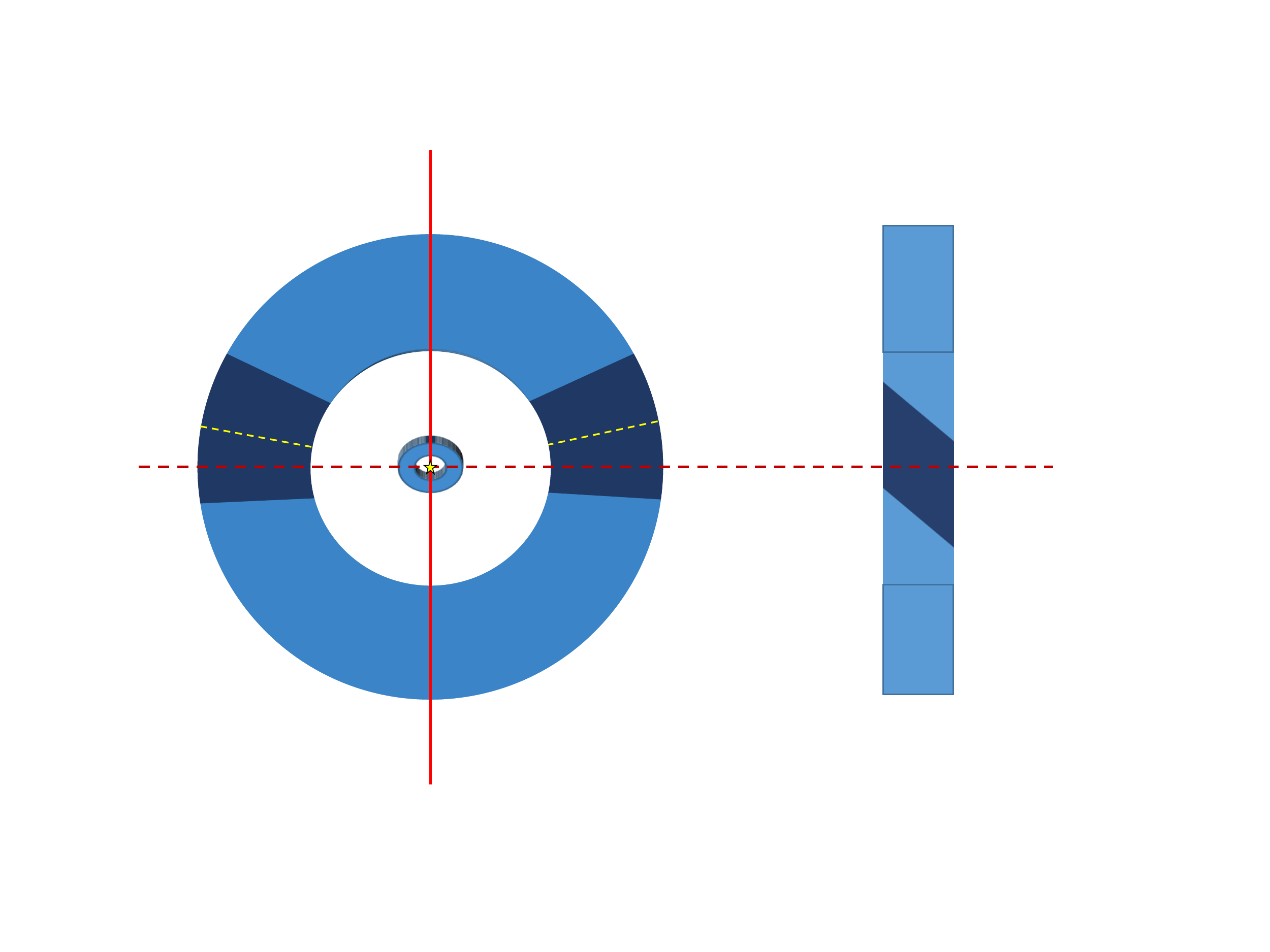}
\caption{Figure showing both a face on schematic of the outer disk (left) and a vertical cross-section of the outer disk (right). The red dashed line represents the major axis of the inner disk and the solid line denotes the vertical slice along which the cross section is produced. The inner disk is tilted with the top edge closer to the observer in this schematic and the dark regions represent the shadows it casts as it intersects the light coming from the central star (represented by the yellow star). In the cross-section we can see how the shadow is shifted vertically in the direction of the tilt of the inner disk. The face-on view of the outer disk shows how the center of these shadows (yellow dashed line) can be offset from the view of the observer. Because of this offset, the separation of the shadows will always be less than 180$\degr$ on the side of the outer disk corresponding to the tilt of the inner disk.}
\label{fig:Disk Cross-section}
\end{figure}

Due to thickness of the outer disk, we find that any $\Delta i \neq \vert90\degr\vert$ will offset the apparent location of the shadows on the outer disk in the direction of the tilt of the inner disk (Figure ~\ref{fig:Disk Cross-section}). The shadows are cast along the major axis of the inner disk and create a shadow, which is not coplanar with the outer disk, on the inner edge of the outer disk. The shadow will therefore be shifted by an amount which depends both on the thickness of the outer disk and the value of $\Delta i$. This suggests the separation of shadows must be less than 180$\degr$ on the side of the outer disk which corresponds to the near side of the inner disk as seen in Figure~\ref{fig:Disk Cross-section}. Examination of the shadow location therefore allows for a simple, effective method of determining the near side of the inner disk. In the case of HD 100453 it is clear that the northern side of the inner disk is the near side because the shadow separation is less than $180\degr$ on that side. In contrast, it is much more difficult to discern the near side of the outer disk via examination of the shadows as this depends on both the tilt of the outer disk and $\Delta i$. In this case, because we know $\Delta i = 45\degr$, the inclination of the outer disk is $\vert25\degr\vert$ from Section 3.3.3, and from the NIR portion of the SED we are not looking along the edge of the inner disk, the SW side of the disk must be the near side.

\section{DISCUSSION} 

\subsection{Dropoff in Spiral Arm Intensity with Wavelength}

We observed a dropoff in spiral arm intensity at longer wavelengths in SPHERE TI imagery, most prevalent in K1 (Figure~\ref{fig:r^2 SPHERE and GPI}), which suggests they are made up of small, compact grains. Though there is a low gas to dust ratio in the disk of HD 100453 \citep{kama16} there is also no detection of a small dust grain dominated tail by HST ACS \citep{collins09}. The lack of a tail, as seen in HD 141569 \citep{konishi16} and young debris disks, suggests the gas to dust is not low enough for radiation pressure to dominate. If gravitational interactions with the M-type companion are the source of the spiral arms, the small grains, which are more tightly coupled to the gas than the large grains, would be pulled more easily with the gas and may lie in a slightly different plane than the rest of the disk. This could explain why the arms are bluer than the rest of the outer disk and according to \citep{dong16} appear to be almost face on. When coupled with the difference in optical depth between the arms and the ring of the outer disk, this suggests that there is a gradient in the particle size distribution of the disk.

\subsection{Rejection of a Coplanar Disk System}

A coplanar inner and outer disk cannot reproduce the shadows seen in the SPHERE and GPI data. From the shadow separation and comparison of model images to data we find a misaligned inner disk with $\Delta i = 45 \pm 10\degr$ and a rotation of the inner disk major axis of -40 $\pm 10\degr$ with respect to the outer disk or 100$ \pm 10\degr$ East of North. This disagrees however with \citet{lazareff16} who find a major axis PA for the inner edge of the inner disk of 81$\pm 1\degr$ East of North. This measurement was found using simplified ellipse and ring models of the inner disk based on VLTI/PIONIER H-band data taken from 2012 December 19 to 2013 February 20.

\subsection{Investigation into Shadow Location Over Time}

At this time there has been no observed change in the shadow pattern over the $\sim11.5$ months spanned by the SPHERE data taken by \citet{wagner15} and \citet{benisty17}. There is the possibility however, of a 19$\degr$ change in major axis position angle between the \citet{lazareff16} measurement and the \citet{wagner15} measurement which spans $\sim25-27$ months. If this is a change in the major axis it could be due to precession of the inner disk, orbital motion of material in the inner disk, or the discrepancy could be a warp between the shadowing structure and the inner region of the inner disk. 

Though the cause of the misalignment between the inner and outer disk is unknown, it has been shown in $\beta$ Pic \citep{lagrange10} that a planet can warp the inner disk. If a planet did cause the misalignment of the inner disk in HD 100453 as suggested above, it would likely cause precession of the inner disk, and by extension the shadows. However, this precession would occur on a timescale on the order of $10^3$ times the orbital timescale expected from Newtonian dynamics \citep{rawiraswattana16}. This suggests we should not observe a change in the location of the shadows due to precession for timescales less than a decade. A change of $19 \pm 10\degr$ in $\sim25-27$ months ($\sim16\degr yr^-1$) is too large to be accounted for with precession and is therefore not the cause of the discrepancy.

The orbital radius of an object with this orbital motion would be $\sim10$ au. This is much larger than the proposed outer radius of the inner disk and just outside the region excluded by speckle residuals in GPI data. The large radius and lack of any disk detection in GPI data suggest this is also likely not the cause of the discrepancy between major axis PAs.

This leaves us with a warp between the inner region of the inner disk (traced by VLTI/PIONIER) and a shadowing region further out. Though this cannot be ruled out using existing GPI or SPHERE data, N-band measurements using VLTI/MATISSE would allow for the angular resolution necessary to detect any difference in the inner and outer portions of the inner disk.

\subsection{Frequency of Transitional Disks with Misaligned Inner Disks}

To date, three of the fourteen, $\sim21\%$, transitional disks of with published images have exhibited these shadows which suggests they are not uncommon. The origin of the shadows however, is still under debate. Suggested causes include: stellar magnetic activity \citep{pffeiferanddong04}, perturbation by the binary companion \citep{martin14}, an unseen planet on an inclined orbit \citep{mouillet97}, etc. However magnetic field strengths in Herbig stars like HD 100453A are small, $\sim$100G \citep{hubrig15} compared to the magnetic fields required for disk misalignment $\sim 10^3$G \citep{pffeiferanddong04}. A close binary system can cause a misaligned inner disk however \citet{collins09} did not observe a strong X-ray source at the location of HD 100453A which would be indicative of another M-type companion, and brighter stars would be visible in the SED. Alternatively a wide binary system, where the companion lies outside of circumstellar disk (as in the case of HD 100453), can also cause misaligned disks. However, this causes large outer disk eccentricity at higher misinclinations \citep{martin14} which is not seen here. The possibility of an unseen planet however, makes HD 100453 a promising candidate for future planet searches \citep{gratiaandfabrycky16}. To date however, there have been no radial velocity measurements taken for the star with the intent of planet detection. In addition this system resembles $\beta$~Pic in having a two belt architecture and a misaligned inner disk, indicating that similar structure in young debris disks is likely inherited from the transitional disk phase.

\section{Conclusions}

We have carried out GPI PI imaging of HD 100453 in Y-, J-, and K1-bands, and have examined the circumstellar disk using this data, the SED, and archival SPHERE TI imagery. With the help of the Monte Carlo Radiative tranfer code HOCHUNK3D, we have generated models which allowed us to probe the inner and outer disk morphology and make several testable predictions about the structure of the disk. Our conclusions are:

\begin{itemize}

\item The circumstellar disk of HD 100453 contains an inner disk which SED fitting suggests extends from 0.13 - 1 au, followed by a large radial gap (1 - 18 au), and an outer disk (18 - 39 au).

\item The outer disk is red in TI imagery, with \(\frac{fdisk}{ftotal}\) = $0.018 \pm 0.002$ in J-band and \(\frac{fdisk}{ftotal}\) = $0.025 \pm 0.002$ in H-band. This is similar in disk color to HR 4796, HD 169142, and 142527 \citep{schneider09,fukagawa10}, and is inconsistent with ISM-like grains at the disk surface. It's low gas to dust ratio \citep{kama16} point to HD 100453 evolving into a young debris disk. 

\item Both TI and PI images exhibit azimuthally-localized dark features at similar PAs. The size of the features is sufficiently large that any single body clearing them would be detected as a bright source \citep{janson12}. Given the sharpness and narrow size of the features and their consistent appearance in PI and TI light we also exclude inhomogeneities in grain properties. We therefore suggest that they are shadows cast by an optically thick, misaligned inner disk.

\item An outer disk inclination of $5\degr$ \citep{dong16} from face on is not supported by the data. In contrast we measure an inclination of $25 \pm 10\degr$, in broad agreement with the 34$\degr$ inclination reported in \citet{wagner15}. 

\item The degree of misalignment between the inner and outer disk can be determined using the separation of the shadows. We find this separation is best reproduced with a $\Delta i = 45 \pm 10\degr$ which also gives us the best fit to the NIR region of the SED.

\item Examination of the shadows can constrain the thickness of the inner disk. As the inner disk thickness increases the width of the shadows it causes also increases. 

\item There is a difference in the color of the spiral arms and the rest of the outer disk. This suggests that they are likely made of small grains which couple to gas more easily and are separated from the outer disk through interaction with the M-type companion.

\item There is  a discrepancy of $\sim19\degr$ between the major axis PA for the inner disk found by \citet{lazareff16} and the one found in our study. Because of the long timescale required for precession and the lack of a visible blob at 10 au, we suggest that the cause of this discrepancy is a warp between the inner portion of the inner disk and an outer shadowing portion.

\end{itemize}

We have shown that shadow morphology and location constrains both the thickness and orientation of the inner disk as well as the flaring of the outer disk. This offers an independent means of measuring structure in transitional disks. In addition the possibility of differences between the inner and outer edges of the inner disk is intriguing and should be investigated further. Predictions of the shadowing model will be tested for this and other Herbig Ae/Be stars with VLTI/MATISSE and other interferometric instruments in the future.

\acknowledgements{Based in part on data obtained at the Gemini Observatory via the time exchange program between Gemini and the Subaru Telescope (GS-2015A-C-1). The Gemini Observatory is operated by the Association of Universities for Research in Astronomy, Inc., under a cooperative agreement with the NSF on behalf of the Gemini partnership: the National Science Foundation (United States), the National Research Council (Canada), CONICYT (Chile), Ministerio de Ciencia, Tecnolog\'{i}a e Innovaci\'{o}n Productiva (Argentina), and Minist\'{e}rio da Ci\^{e}ncia, Tecnologia e Inova\c{c}\~{a}o (Brazil). MT is partly supported by JSPS KAKENHI 2680016. CAG is supported under NASA Origins of Solar Systems Funding via NNG16PX39P. Y.H. is supported by Jet Propulsion Laboratory, California Institute of Technology under a contract from NASA. MS is supported by NASA Exoplanet Research Program NNX16AJ75G. J.K. acknowledges support from Philip Leverhulme Prize (PLP-2013-110, PI: Stefan Kraus). S.K. acknowledges support from an ERC Starting Grant (Grant Agreement No. 639889). We also thank the referee for their comments and suggestions which added clarity to this paper.}

\pagebreak

\begin{thebibliography}{}


\bibitem[Andrews et al.(2016)]{andrews16} Andrews, S., M., Wilner, D. J., Zhu, Z., et al. 2016, ApJL, 820, 2

\bibitem[Apai et al.(2015)]{apai15} Apai, D., Schneider, G., Grady, C., et al. 2015, ApJ, 800, 136

\bibitem[Benisty et al.(2017)]{benisty17} Bensity, M., Stolker, T., Pohl, A., et al. 2016, A\&A, 597, A42

\bibitem[Birnstiel et al.(2012)]{birnstiel12} Birnstiel, T., Andrews, S. M., \& Ercolano, B. A\&A 544A, 79B

\bibitem[Brott \& Hauschildt(2005)]{brott05} Brott, I., \& Hauschildt, P. H. 2005, in The Three-Dimensional Universe with Gaia, ed. C. Turon, K. S. O’Flaherty, \& M. A. C. Perryman (ESA SP-576; Noordwijk: ESA), 565


\bibitem[Chen et al.(2006)]{chen06} Chen, X. P., Henning, T., van Boekel, R., \& Grady, C. A. 2006 A\&A, 445, 331

\bibitem[Chiang and Goldreich(1997)]{chiang97} Chiang, E. I., \& Goldreich, P. 1997, ApJ, 490, 368

\bibitem[Collins et al.(2009)]{collins09} Collins, K. A., Grady, C. A., \& Hamaguchi, K., et al. 2009, ApJ, 697, 557

\bibitem[Cutrie et al.(2003)]{cutrie03} Cutrie, R. M., Skrutskie, M. F., van Dyk, S., et al. 2003, The IRSA 2MASS All-Sky Point Source Catalog (NASA/IPAC Infrared Science Archive) 

\bibitem[Dominik et al.(2003)]{dominik03} Dominik, C., Dullemond, C. P., Waters, L. B. F. M., \& Walch, S. 2003, A\&A, 398, 607

\bibitem[Dullemond \& Dominik(2004a)]{dullemondanddominik04a} Dullemond, C. P. \& Dominik, C. 2004a, A\&A, 417, 159

\bibitem[Dullemond \& Dominik(2004b)]{dullemondanddominik04b} Dullemond, C. P. \& Dominik, C. 2004b, A\&A, 421, 1075


\bibitem[Dong et al.(2016)]{dong16} Dong, R. Zhu, Z., Fung, J., Rafikov, R., Chiang, E., Wagner, \& K., 2015, ApJL, 815, 1

\bibitem[Fukagawa et al.(2010)]{fukagawa10} Fukagawa, M., Tamura, M., Itoh, Y., et al. 2010, PASJ, 62, 347-370 

\bibitem[Girard et al.(2016)]{gpimanual} Girard, J., Wahhaj, Z., Vigan, A., et al. 2016, PDM-ESO-254263 VLT-MAN-SPH-14690-0430, Very Large Telescope SPHERE User Manual, https://www.eso.org/sci/facilities/paranal/
instruments/sphere/doc.html

\bibitem[Gaia Collaboration et al.(2016a)]{gaia16} Gaia Collaboration et al. (2016a): Summary description of Gaia DR1.

\bibitem[Gratia and Fabrycky(2016)]{gratiaandfabrycky16} Gratia, P. and Fabrycky, D., 2016, arXiv, 1607.08630

\bibitem[Greaves et al.(2014)]{greaves14} Greaves, J. S., Kennedy, G. M., Thureau N. et al., 2014, MNRAS, 438, L31

\bibitem[Guimar\~aes et al.(2006)]{guimaraes06} Guimar\~aes, M. M., Alencar, S.H.P., Corradi, W.J.B., \& S. L. A. Vieira. 2006, A\&A, 457, 581

\bibitem[Hartmann et al. (1998)]{hartmann98} Hartmann, L., Calvet, N., Gullbring, E., \& D’Alessio, P. 1998, ApJ, 495, 385

\bibitem[Hamilton and Burns (1992)]{hamiltonandburns92} Hamilton, D. P., and Burns, J. B. 1992, Icarus, 96, 43

\bibitem[Hubrig et al.(2015)]{hubrig15} Hubrig, S., Carroll, T. A., Sch\"oller, M., \& Ilyin, I. 2015, MNRAs, 449, 118

\bibitem[Janson et al.(2012)]{janson12} Janson, M., Jayawardhana, R., Girard, J., et al., 2012, ApJL, 758, L2

\bibitem[Kama et al.(2016)]{kama16} Kama, M., Bruderer, S., Carney, M. et al., 2016, A\&A, 588, A108 

\bibitem[Khalafinejad et al.(2016)]{khalafinejad16} Khalafinejad, S., Maaskant, K., M., Mari\~nas, N., \& Tielens, A.G.G.M., 2016, A\&A, 587, A62

\bibitem[Konishi et al.(2016)]{konishi16} Konishi, M., Grady, C. A., Schneider, G., et al. 2016, ApJL, 818, 2

\bibitem[Lazareff et al.(2016)]{lazareff16} Lazareff, B., Berger, J.-P., Kluska, J., et al. 2016, A\&A, submitted

\bibitem[Lagrange et al.(2010)]{lagrange10} Lagrange, A. M., Bonnefoy, M., Chauvin, G., et al. 2010, Science, 329, 5987

\bibitem[Lucy (1999)]{lucy99} Lucy, L. B. 1999, A\&A, 344, 282

\bibitem[Maaskant et al.(2013)]{maaskant13} Maaskant, K. M., Honda, M., \& Waters, L. B. F. M., et al. 2013, A\&A, 555, A64

\bibitem[Macintosh et al.(2014)]{macintosh14} Macintosh, B., Graham, J. R., Ingraham, P., et al. 2014, PNAS, 111, 35

\bibitem[Maire et al.(2010)]{maire10} Maire, J., Perrin, M. D., Doyon, R. et al., 2010, Proceedings of the SPIE, 7735, id. 773531

\bibitem[Marino et al.(2015)]{marino15} Marino, S., Perez, S., \& Casassus, S., 2015, ApJL, 198, 2 

\bibitem[Martin et al.(2014)]{martin14} Martin, R., Nixon, C., Lubow, S., et al., 2014, ApJL, 792, L33

\bibitem[Meeus et al.(2002a)]{meeus02a} Meeus, G., Bouwman, J., \& Dominik, C., et al. 2002 A\&A, 392, 1039-1046

\bibitem[Meeus et al.(2002b)]{meeus02b} Meeus, G., Bouwman, J., \& Dominik, C., et al. 2002 A\&A, 402, 767

\bibitem[Menu et al.(2015)]{menu15} Menu, J., van Boekel, R., Henning, Th., et al. A\&A, 581, A107

\bibitem[Mouillet et al.(1997)]{mouillet97} Mouillet, D., Larwood, J. D., Papaloizou, J. C. B., \& Lagrange, A. M. 1997, MNRAS, 292, 896 

\bibitem[Mulders et al.(2013)]{mulders13} Mulders, G. D., Min, M., \& Dominic C., 2013, A\&A, 549, A112

\bibitem[Pascual et al.(2016)]{pascual16} Pascual, N., Montesinos, B., Meeus, G., et al., 2015, A\&A, 585, A6

\bibitem[Perrin et al.(2014)]{perrin14} Perrin, M. D., Maire, J., Ingraham, P. et al., 2014, Proceedings of the SPIE, 9147, id. 91473J

\bibitem[Pffeifer and Dong(2004)]{pffeiferanddong04} Pffeifer, H. P. \& Dong, L. 2004, ApJ, 604, 766

\bibitem[Rawiraswattana et al.(2016)]{rawiraswattana16} Rawiraswattana, K., Hubber, \& D. A., Goodwin, S. P., MNRAS, 460, 4 

\bibitem[Schneider et al.(2009)]{schneider09} Schneider, G., Weinberger, A. J., Becklin, E. E., et al., 2009, ApJ, 137, 53-61

\bibitem[Slettebak et al.(1966)]{slettebak66} Slettebak A., ApJ, 145, 126S

\bibitem[Stolker et al.(2016)]{stolker16} Stolker, T., Dominik, C., Avenhaus, H., et al., 2016, A\&A, http://arxiv.org/abs/1603.00481 

\bibitem[Wagner et al.(2015)]{wagner15} Wagner, K., Apai, D., Kasper, M., \& Robberto, M., 2015, ApJL, 813, L2

\bibitem[Whitney et al.(2003a)]{whitney03a} Whitney, B.A., Wood,K., Bjorkman, J.E., \& Wolff, M., 2003, ApJ, 591, 1049



\bibitem[Whitney et al.(2013)]{whitney13} Whitney, B.A., Robitaille, T.P., Bjorkman, J.E., Dong, R., Wolff, M.J., Wood, K., \& Honor, J., 2013

\bibitem[Wood et al.(2002)]{wood02} Wood, K., Wolff, M.J., Bjorkman, J.E., \& Whitney, B., 2002, ApJ, 564, 887

\end{thebibliography}
\end{document}